\documentclass[12pt]{article}
\usepackage{amssymb}
\begin{document}



\def\a{\alpha}
\def\b{\beta}
\def\d{\delta}
\def\e{\epsilon}
\def\g{\gamma}
\def\h{\mathfrak{h}}
\def\k{\kappa}
\def\l{\lambda}
\def\o{\omega}
\def\p{\wp}
\def\r{\rho}
\def\t{\theta}
\def\s{\sigma}
\def\z{\zeta}
\def\x{\xi}
 \def\A{{\cal{A}}}
 \def\B{{\cal{B}}}
 \def\C{{\cal{C}}}
 \def\D{{\cal{D}}}
\def\G{\Gamma}
\def\K{{\cal{K}}}
\def\O{\Omega}
\def\R{\bar{R}}
\def\T{{\cal{T}}}
\def\L{\Lambda}
\def\f{E_{\tau,\eta}(sl_2)}
\def\E{E_{\tau,\eta}(sl_n)}
\def\Zb{\mathbb{Z}}
\def\Cb{\mathbb{C}}

\def\R{\overline{R}}

\def\beq{\begin{equation}}
\def\eeq{\end{equation}}
\def\bea{\begin{eqnarray}}
\def\eea{\end{eqnarray}}
\def\ba{\begin{array}}
\def\ea{\end{array}}
\def\no{\nonumber}
\def\le{\langle}
\def\re{\rangle}
\def\lt{\left}
\def\rt{\right}

\newtheorem{Theorem}{Theorem}
\newtheorem{Definition}{Definition}
\newtheorem{Proposition}{Proposition}
\newtheorem{Lemma}{Lemma}
\newtheorem{Corollary}{Corollary}
\newcommand{\proof}[1]{{\bf Proof. }
        #1\begin{flushright}$\Box$\end{flushright}}

\baselineskip=20pt

\newfont{\elevenmib}{cmmib10 scaled\magstep1}
\newcommand{\preprint}{
   \begin{flushleft}
   \end{flushleft}\vspace{-1.3cm}
   \begin{flushright}\normalsize
   \end{flushright}}
\newcommand{\Title}[1]{{\baselineskip=26pt
   \begin{center} \Large \bf #1 \\ \ \\ \end{center}}}
\newcommand{\Author}{\begin{center}
   \large \bf

Xi Chen${}^{a}$,~Wen-Li Yang${}^{a}$,~Xiang-Mao Ding${}^{b}$,~Jun Feng${}^{a}$,\\San-Min Ke${}^{c}$,~Ke Wu${}^d$
~and~Yao-Zhong Zhang${}^e$

\end{center}}
\newcommand{\Address}{\begin{center}

${}^a$ Institute of Modern Physics, Northwest University,
       Xian 710069, P.R. China\\
${}^b$ Institute of Applied Mathematics, Academy of Mathematics and Systems Sciences, Chinese Academy of Sciences,
       P.O.Box 2734, Beijing 100080, P.R. China\\
${}^c$ College of Science, Chang'an University, Xian 710064, P.R. China\\
${}^d$ School of Mathematical Science, Capital Normal University,
     Beijing 100037, P.R. China \\
${}^e$ School of Mathematics and Physics, The University of Queensland, Brisbane,
       QLD 4072, Australia

   \end{center}}
\newcommand{\Accepted}[1]{\begin{center}
   {\large \sf #1}\\ \vspace{1mm}{\small \sf Accepted for Publication}
   \end{center}}

\preprint
\thispagestyle{empty}
\bigskip\bigskip\bigskip

\Title{
Free field realization of  the exceptional
current superalgebra $\widehat{D(2,1;\a)}_k$
} \Author

\Address
\vspace{1cm}

\begin{abstract}
The free-field representations of the  $D(2,1;\a)$ current superalgebra and the
corresponding  energy-momentum tensor are constructed. The related screening
currents of the first kind are also presented.

\vspace{1truecm} \noindent {\it PACS:} 11.25.Hf; 02.20.Tw

\end{abstract}
\newpage
\section{Introduction}
\label{intro} \setcounter{equation}{0}

The interest in two-dimensional non-linear sigma models with
supermanifold as target space has drastically increased over the last ten years
due to their applications in superstring theory \cite{Ber99,Bers99,Sch06,Que07},
logarithmic conformal field theory (CFT)  and condensed
matter physics \cite{Efe83,Ber95,Mud96,Maa97,Bas00, Gur00, Lud00, Bha01}.
Among them, the sigma models associated with superalgebras
$psl(n|n)$ and $osp(2n+2|2n)$ stand out as an important class. These supergroups have vanishing dual
Coexter number. As a result, sigma models based on the supergroups have vanishing one-loop
beta function and are thus expected to be conformal invariant even without adding the Wess-Zumino
terms \cite{Bers99}.  Recent studies show that the superalgebra $D(2,1;\a)$,
which is the one-parameter deformation of superalgebra $osp(4|2)$ and has a vanishing dual
Coexter number,  has played an important role in describing the origin of the Yangian symmetry of Ads/CFT \cite{Mat08}.
Here we shall study the Wess-Zumino-Novikov-Witten (WZNW) model associated with $D(2,1;\a)$.

Free field realization \cite{Fra97,Fei90,Bou90,Ito90,Fre94,Ras98,Din03,Zha05,Yan06,Yan08,Yan09}
has been proved to be a powerful method in analyzing CFTs such
as WZNW models. In this letter, motivated by its potential applications,  we
obtain the free-field realization of the  current superalgebra underlying the $D(2,1;\a)$
WZNW model.


\section{$D(2,1;\a)$ current algebra}
\label{XXZ} \setcounter{equation}{0}

Let us start with some basic notation of the $D(2,1;\a)$ current
algebra, i.e. the $\widehat{D(2,1;\a)}$ affine superalgebra
\cite{Fra96}. The exceptional Lie superalgebra $D(2,1;\a)$ forms a
one-parameter (i.e. $\a$) family of superalgebras of rank 3 and
dimension 17. The bosonic (or even) part  is a direct
sum of three $su(2)$ and the fermionic (or odd) part is a spinorial
representation $(2,2,2)$ of the even part.
Let $\Pi:=\{\a_1,\,\a_2,\,\a_3\}$ be the simple roots with $\a_1$ being fermionic
and $\a_2$, $\a_3$ being bosonic. The Cartan matrix $a_{ij}$
is given by
\bea
  a=\lt(\begin{array}{ccc}0&1&\a\\-1&2&0\\-1&0&2\end{array}\rt).\label{Cartan}
\eea
The positive even roots $\Delta^{+}_{0}$ and the positive odd roots $\Delta^{+}_{1}$ are then given respectively by
\bea
  \Delta^{+}_{0}=\lt\{\a_2,\,\a_3,\,2\a_1+\a_2+\a_3\rt\},\quad\quad
  \Delta^{+}_{1}=\lt\{\a_1,\,\a_1+\a_2,\,\a_1+\a_3,\,\a_1+\a_2+\a_3\rt\},
\eea the set of all positive roots $\Delta^+$ is a union of the positive even and odd roots, namely,
\bea
   \Delta^+ =\Delta^{+}_{0}\bigcup\Delta^{+}_{1}.
\eea
Associated with each positive root $\delta$, there is a raising operator $E_{\d}$, a lowering operator
$F_{\d}\equiv E_{-\d}$ and a Cartan generator $H_{\d}$. These operators have definite $\Zb_2$-gradings:
\bea
  [H_{\d}]=0,\quad\quad [E_{\d}]=[F_{\d}]=\lt\{\begin{array}{ll}
  0,&\d\in \Delta^{+}_{0},\\[8pt]
  1,&\d\in \Delta^{+}_{1}.\end{array}\rt.
\eea For any two homogenous elements (i.e.
elements with definite $\Zb_2$-gradings) $a,b\in D(2,1;\a)$, the Lie bracket is defined by
\bea
 \lt[a,\,b\rt]=ab-(-)^{[a][b]}ba,
\eea this (anti)commutator extends to inhomogenous elements through linearity. Then the
commutation relations of $D(2,1;\a)$ are \cite{Fra96}
\footnote{Here we only give the relevant relations for our purpose, the others can be found in \cite{Fra96}}
\bea
  &&[E_{\a_i},\, F_{\a_j}]=\d_{ij}\,H_{\a_i},\quad\quad [H_{\a_i},\,H_{\a_j}]=0,\quad\quad i,j=1,2,3,
     \label{commutation relation-1}\\
  &&[H_{\a_i},\, E_{\a_j}]=a_{ij}E_{\a_j},\quad\quad [H_{\a_i},\, F_{\a_j}]=-a_{ij}F_{\a_j},
     ,\quad\quad i,j=1,2,3,\no\\
  &&[E_{\a_1},\,E_{\a_2}]=-E_{\a_1+\a_2},\quad\quad [E_{\a_1},\,E_{\a_3}]=-E_{\a_1+\a_3},\no\\
  &&[E_{\a_1},\,E_{\a_1+\a_2+\a_3}]=-(1+\a)E_{2\a_1+\a_2+\a_3}, \no\\
  &&[E_{\a_2},\,E_{\a_1+\a_3}]=[E_{\a_3},\,E_{\a_1+\a_2}]=E_{\a_1+\a_2+\a_3},\no\\
  &&[E_{\a_1+\a_2},\,E_{\a_1+\a_3}]=(1+\a)E_{2\a_1+\a_2+\a_3},\no\\
  &&[F_{\a_1},\,E_{2\a_1+\a_2+\a_3}]=-E_{\a_1+\a_2+\a_3},\quad\quad
    [F_{\a_1},\,E_{\a_1+\a_2}]=-E_{\a_2},\no\\
  &&[F_{\a_1},\,E_{\a_1+\a_3}]=-\a E_{\a_3},\quad\quad [F_{\a_2},\,E_{\a_1+\a_2+\a_3}]=E_{\a_1+\a_3},
     \no\\
  &&[F_{\a_2},\,E_{\a_1+\a_2}]=E_{\a_1},\quad\quad [F_{\a_3},\,E_{\a_1+\a_3}]=E_{\a_1},\no\\
  &&[F_{\a_3},\,E_{\a_1+\a_2+\a_3}]=E_{\a_1+\a_2}.\label{commutation relation-2}
  \eea

One may introduce a nondegenerate and invariant supersymmetric metric or bilinear form,
denoted by $(X,Y)$ for any $X,Y \in D(2,1;\a)$, as follows:
\bea
 &&(E_{\a_1},\, F_{\a_1})=-(F_{\a_1},\, E_{\a_1})=-1,\,\,
   (E_{\a_1+\a_2},\, F_{\a_1+\a_2})=-(F_{\a_1+\a_2},\, E_{\a_1+\a_2})=1,\label{Inner-product-1}\\
 &&(E_{\a_1+\a_3},\, F_{\a_1+\a_3})=-(F_{\a_1+\a_3},\, E_{\a_1+\a_3})=1,\no\\
 &&(E_{\a_1+\a_2+\a_3},\, F_{\a_1+\a_2+\a_3})=-(F_{\a_1+\a_2+\a_3},\, E_{\a_1+\a_2+\a_3})=-1,\no\\
 &&(E_{2\a_1+\a_2+\a_3},\, F_{2\a_1+\a_2+\a_3})=(F_{2\a_1+\a_2+\a_3},\, E_{2\a_1+\a_2+\a_3})
    =-\frac{1}{1+\a},\no\\[6pt]
 &&(E_{\a_2},\, F_{\a_2})=(F_{\a_2},\, E_{\a_2})=1,\,\,
   (E_{\a_3},\, F_{\a_3})=(F_{\a_3},\, E_{\a_3})=\frac{1}{\a},\no\\[6pt]
 &&(H_{\a_1},\,H_{\a_1})=0,\,\,(H_{\a_1},\,H_{\a_2})=(H_{\a_2},\,H_{\a_1})=1,
   \,\,(H_{\a_1},\,H_{\a_3})=(H_{\a_3},\,H_{\a_1})=1,\no\\[6pt]
 &&(H_{\a_2},\,H_{\a_2})=2,\quad\quad(H_{\a_2},\,H_{\a_3})=(H_{\a_3},\,H_{\a_2})=0,
    \quad\quad (H_{\a_3},\,H_{\a_3})=\frac{2}{\a}.\label{Inner-product-2}
\eea All the other inner products are zero. With the help of the above inner product we can
construct  the corresponding second-order Casimir element
\bea
 C_2&=&(E_{\a_1}F_{\a_1}-F_{\a_1}E_{\a_1})-(E_{\a_1+\a_2}F_{\a_1+\a_2}-F_{\a_1+\a_2}E_{\a_1+\a_2})
        -(E_{\a_1+\a_3}F_{\a_1+\a_3}-F_{\a_1+\a_3}E_{\a_1+\a_3})\no\\
    &&+(E_{\a_1+\a_2+\a_3}F_{\a_1+\a_2+\a_3}-F_{\a_1+\a_2+\a_3}E_{\a_1+\a_2+\a_3})
      +(E_{\a_2}F_{\a_2}+F_{\a_2}E_{\a_2})\no\\
    &&-(1+\a)(E_{2\a_1+\a_2+\a_3}F_{2\a_1+\a_2+\a_3}+F_{2\a_1+\a_2+\a_3}E_{2\a_1+\a_2+\a_3})
      +\a(E_{\a_3}F_{\a_3}+F_{\a_3}E_{\a_3})\no\\[6pt]
    &&\frac{1}{1+\a}\left\{-2H^2_{a_1}+2H_{\a_1}H_{\a_2}+2\a H_{\a_1}H_{\a_3}
      +\frac{\a}{2}H_{\a_2}^2-\a H_{\a_2}H_{\a_3}+\frac{\a}{2}H_{\a_3}^2
    \right\}.\label{Casimir}
\eea

Let $X(z)$ be current associated with generator $X\in D(2,1;\a)$. For an example,
denote the currents corresponding to $E_{\d},\, H_{\d},\,F_{\d}$ ($\forall \d\in \Delta^+$) by
$E_{\d}(z),\, H_{\d}(z),\,F_{\d}(z)$ respectively.
The $D(2,1;\a)$ current algebra is generated by a set of currents
$\{X(z)|\,X\in D(2,1;\a)\}$, which obey the
following OPEs \cite{Fra97},
\bea
  X(z)\,Y(w)=k\frac{\lt(X,Y\rt)}{(z-w)^2}+\frac{[X,\,Y]\,(w)}{(z-w)}, \label{OPE}
\eea where $k$ is the level (or central charge) of the current algebra.

\section{Free field realization}
 \label{FFR} \setcounter{equation}{0}
To obtain a free field realization of the $D(2,1;\a)$ currents, one needs
\cite{Fei90,Bou90,Ito90,Fre94,Ras98,Din03,Zha05,Yan06,Yan08,Yan09} firstly to
construct  the differential operator representation  of the corresponding  finite-dimensional
superalgebra $D(2,1;\a)$. A particular ordering of all positive roots of
the superalgebra is crucial \cite{Yan09} in getting the explicit expression of the
differential operator representation.  For the $D(2,1;\a)$ case, the ordering
of the positive roots is given by
\bea
 \a_3,\,\a_2,\,\a_1+\a_2+\a_3,\,\a_1+\a_3,\,2\a_1+\a_2+\a_3,\,\a_1+\a_2,\,\a_1.
\eea Such an ordering allows us to construct the explicit expression of the differential
operator representation of $D(2,1;\a)$ by the method developed in \cite{Yan09}.
With the help of the differential representation we can obtain the
free field realization of the $D(2,1;\a)$
current algebra in terms of $3$ bosonic $\b$-$\g$ pairs
($(\b_{\a_2}(z),\,\g_{\a_2}(z))$, $(\b_{\a_3}(z),\,\g_{\a_3}(z))$ and
$(\b_{2\a_1+\a_2+\a_3}(z),\,\g_{2\a_1+\a_2+\a_3}(z)))$, $4$ fermionic $b$-$c$ pairs
(($\Psi^{\dagger}_{\a_1}(z),\,\Psi_{\a_1}(z))$, ($\Psi^{\dagger}_{\a_1+\a_2}(z),\,\Psi_{\a_1+\a_2}(z))$,
($\Psi^{\dagger}_{\a_1+\a_3}(z),\,\Psi_{\a_1+\a_3}(z))$ and ($\Psi^{\dagger}_{\a_1+\a_2+\a_3}(z)$, $\,
\Psi_{\a_1+\a_2+\a_3}(z)))$ and
$3$ free scalar fields ($\phi_i(z)$, for $i=1,2,3$). These free
fields obey the following OPEs:
\bea
&& \b_{i}(z)\,\g_{j}(w)=-\g_{j}(z)\,\b_{i}(w)=
   \frac{\d_{ij}}{z-w},\quad\quad i,j\in\Delta^+_{0},\label{OPE-F-1}\\[6pt]
&& \psi_{i}(z)\psi_{j}^{\dagger}(w)=\psi_{j}^{\dagger}(z)
   \psi_{i}(w)=\frac{\d_{ij}}{z-w},\quad\quad i,j\in\Delta^+_1,\\
&& \phi_1(z)\phi_1(w)=-\ln(z-w),\\
&& \phi_i(z)\phi_j(w)=\d_{ij}\ln(z-w),\quad\quad i,j=2,3.
   \label{OPE-F-2}\eea  and the other OPEs are trivial.
Here we present the results for
the currents associated with the simple roots,
\bea
 E_{\a_1}(z)&=&\Psi_{\a_1}(z),\label{Free-field-realization-1}\\
 E_{\a_2}(z)&=&-\Psi^+_{\a_1}(z)\Psi_{\a_1+\a_2}(z)
         -\Psi^+_{\a_1+\a_3}(z)\Psi_{\a_1+\a_2+\a_3}(z)
         +\b_{\a_2}(z),\\
 E_{\a_3}(z)&=&-\Psi^+_{\a_1}(z)\Psi_{\a_1+\a_3}(z)
         -(1+\a)\Psi^+_{\a_1}(z)\Psi^+_{\a_1+\a_2}(z)\b_{2\a_1+\a_2+\a_3}(z)\no\\
         &&-\Psi^+_{\a_1+\a_2}(z)\Psi_{\a_1+\a_2+\a_3}(z)+\b_{\a_3}(z),\\
 H_{\a_1}(z)&=&-\Psi^+_{\a_1+\a_2}(z) \Psi_{\a_1+\a_2}(z)
           -(1+\a)\g_{2\a_1+\a_2+\a_3}(z)\b_{2\a_1+\a_2+\a_3}(z)\no\\
         &&-\a\Psi^+_{\a_1+\a_3}(z)\Psi_{\a_1+\a_3}(z)
           -(1+\a)\Psi^+_{\a_1+\a_2+\a_3}(z)\Psi_{\a_1+\a_2+\a_3}(z)\no\\
         &&-\g_{\a_2}(z) \b_{\a_2}(z)-\a\g_{\a_3}(z)\b_{\a_3}(z) \no\\
         &&+\sqrt{k}\left(\sqrt{\frac{1+\a}{2}}\,\partial\phi_1(z)
           +\frac{\sqrt{2}}{2}\,\partial\phi_2(z)
           +\sqrt{\frac{\a}{2}}\,\partial\phi_3(z)\right),\\
H_{\a_2}(z)&=&\Psi^+_{\a_1}(z)\Psi_{\a_1}(z) -\Psi^+_{\a_1+\a_2}(z)\Psi_{\a_1+\a_2}(z)
           +\Psi^+_{\a_1+\a_3}(z)\Psi_{\a_1+\a_3}(z)\no\\
        && -\Psi^+_{\a_1+\a_2+\a_3}(z)\Psi_{\a_1+\a_2+\a_3}(z)
           -2\g_{\a_2}(z)\b_{\a_2}(z) +\sqrt{2k}\,\partial\phi_2(z),\no\\
H_{\a_3}(z)&=&\Psi^+_{\a_1}(z)\Psi_{\a_1}(z) +\Psi^+_{\a_1+\a_2}(z)\Psi_{\a_1+\a_2}(z)
           -\Psi^+_{\a_1+\a_3}(z)\Psi_{\a_1+\a_3}(z)\no\\
        && -\Psi^+_{\a_1+\a_2+\a_3}(z)\Psi_{\a_1+\a_2+\a_3}(z)
           -2\g_{\a_3}(z)\b_{\a_3}(z) +\sqrt{\frac{2k}{\a}}\,\partial\phi_3(z),\no\\
F_{\a_1}(z)&=&\Psi^+_{\a_1+\a_2}(z)\Psi^+_{\a_1+\a_3}(z)\Psi_{\a_1+\a_2+\a_3}(z)
              -\Psi^+_{\a_1+\a_2}(z)\b_{\a_2}(z)\no\\
           &&+\g_{2\a_1+\a_2+\a_3}(z)\Psi_{\a_1+\a_2+\a_3}(z)
             -\a\Psi^+_{\a_1+\a_3}(z)\b_{\a_3}(z)\no\\
           &&-\Psi^+_{\a_1}(z)\left\{\Psi^+_{\a_1+\a_2}(z)
             \Psi_{\a_1+\a_2}(z)+\a\Psi^+_{\a_1+\a_3}(z)\Psi_{\a_1+\a_3}(z)\right\}\no\\
           &&-\Psi^+_{\a_1}(z)\left\{\g_{\a_2}(z)\b_{\a_2}(z)
             +(1+\a)\g_{2\a_1+\a_2+\a_3}(z)\b_{2\a_1+\a_2+\a_3}(z)
             \right\}\no\\
           &&-\Psi^+_{\a_1}(z)\left\{(1+\a)\Psi^+_{\a_1+\a_2+\a_3}(z)\Psi_{\a_1+\a_2+\a_3}(z)
                        +\a\g_{\a_3}(z)\b_{\a_3}(z)
            \right\} \no\\
           &&+\sqrt{k}\,\Psi^+_{\a_1}(z)\left\{\sqrt{\frac{1+\a}{2}}\,\partial\phi_1(z)
             +\frac{\sqrt{2}}{2}\,\partial\phi_2(z)
             +\sqrt{\frac{\a}{2}}\,\partial\phi_3(z)
           \right\}\no\\
          &&-k\,\partial\Psi^+_{\a_1}(z),\\
F_{\a_2}(z)&=&-\Psi^+_{\a_1+\a_2}(z)\Psi_{\a_1}(z)-\Psi^+_{\a_1+\a_2+\a_3}(z)\Psi_{\a_1+\a_3}(z)
              -\g^2_{\a_2}(z)\b_{\a_2}(z)\no\\
           &&+\sqrt{2k}\,\g_{\a_2}(z)\,\partial\phi_2(z)+(k-2)\,\partial\g_{\a_2}(z),\\
F_{\a_3}(z)&=&(1+\a)\Psi^+_{\a_1+\a_2+\a_3}(z)
              \Psi^+_{\a_1+\a_3}(z)\b_{2\a_1+\a_2+\a_3}(z)\no\\
           &&-\Psi^+_{\a_1+\a_3}(z)\Psi_{\a_1}(z)-\Psi^+_{\a_1+\a_2+\a_3}(z)\Psi_{\a_1+\a_2}(z)-\g^2_{\a_3}(z)\b_{\a_3}(z)
             \no\\
           &&+\sqrt{\frac{2k}{\a}}\,\g_{\a_3}(z)\,\partial\phi_3(z)+(\frac{k}{\a}-2)\,\partial\g_{\a_3}(z).
           \label{Free-field-realization-2}
\eea
Here
normal ordering of the free field expressions is implied. The
free field realization for currents associated with the
non-simple roots can be obtained from the OPEs of the simple ones.
It is straightforward
to check that (\ref{Free-field-realization-1})-(\ref{Free-field-realization-2})
satisfy the OPEs of the $D(2,1;\a)$ current algebra given in the last section.


\section{Energy-momentum tensor}
\label{EMT} \setcounter{equation}{0}

It is well-known that the  energy-momentum tensor associated with a current algebra  can be
obtained by means of the Sugawara's construction.
For the present case,  the Sugawara tensor corresponding to the quadratic
Casimir $C_2$ (\ref{Casimir}) is
\bea
T(z)&=&\frac{1}{2k}\left\{(E_{\a_1}(z)F_{\a_1}(z)-F_{\a_1}(z)E_{\a_1}(z))
       -(E_{\a_1+\a_2}(z)F_{\a_1+\a_2}(z)-F_{\a_1+\a_2}(z)E_{\a_1+\a_2}(z))
       \right.\no\\
    &&\quad\quad -(E_{\a_1+\a_3}(z)F_{\a_1+\a_3}(z)-F_{\a_1+\a_3}(z)E_{\a_1+\a_3}(z))
    \no\\
    &&\quad\quad +(E_{\a_1+\a_2+\a_3}(z)F_{\a_1+\a_2+\a_3}(z)-F_{\a_1+\a_2+\a_3}(z)E_{\a_1+\a_2+\a_3}(z))
      \no\\
    &&\quad\quad -(1+\a)(E_{2\a_1+\a_2+\a_3}(z)F_{2\a_1+\a_2+\a_3}(z)+F_{2\a_1+\a_2+\a_3}(z)E_{2\a_1+\a_2+\a_3}(z))
       \no\\
    &&\quad\quad +(E_{\a_2}(z)F_{\a_2}(z)+F_{\a_2}(z)E_{\a_2}(z))+\a(E_{\a_3}(z)F_{\a_3}(z)+F_{\a_3}(z)E_{\a_3}(z))\no\\
    &&\quad\quad +\frac{1}{1+\a}\left\{-2H_{a_1}(z)H_{a_1}(z)+2H_{\a_1}(z)H_{\a_2}(z)+2\a H_{\a_1}(z)H_{\a_3}(z)
      \right\}\no\\[6pt]
    &&\quad\quad \left.+\frac{1}{1+\a}\left\{\frac{\a}{2}H_{\a_2}(z)H_{\a_2}(z)
      -\a H_{\a_2}(z)H_{\a_3}(z)+\frac{\a}{2}H_{\a_3}(z)H_{\a_3}(z)\right\}
      \right\},
\eea here normal ordering of the free field expressions is implied. The above expression can be written in terms of
the $\b\g$ and $b-c$ pairs, and the scalar fields. After tedious calculation, we obtain
\bea
T(z)&=&-\frac{1}{2}\partial\phi_1(z)\partial\phi_1(z)
       -\sqrt{\frac{1+\a}{2k}}\,\partial^2\phi_1(z)\no\\
    &&+\frac{1}{2}\partial\phi_2(z)\partial\phi_2(z)
       -\sqrt{\frac{1}{2k}}\,\partial^2\phi_2(z)\no\\
    &&+\frac{1}{2}\partial\phi_3(z)\partial\phi_3(z)
       -\sqrt{\frac{\a}{2k}}\,\partial^2\phi_3(z)\no\\
    &&+\b_{\a_2}(z)\partial\g_{\a_2}(z)+\b_{\a_3}(z)\partial\g_{\a_3}(z)
       +\b_{2\a_1+\a_2+\a_3}(z)\partial\g_{2\a_1+\a_2+\a_3}(z)\no\\
    &&+\partial\Psi^+_{\a_1}(z)\,\Psi_{\a_1}(z)+\partial\Psi^+_{\a_1+\a_2}(z)\,\Psi_{\a_1+\a_2}(z)\no\\
    &&+\partial\Psi^+_{\a_1+\a_3}(z)\,\Psi_{\a_1+\a_3}(z)
      +\partial\Psi^+_{\a_1+\a_2+\a_3}(z)\,\Psi_{\a_1+\a_2+\a_3}(z),\label{Tensor}\eea
where the normal ordering of the free field expressions is implicit.  We find
\bea
T(z)T(w)=\frac{c/2}{(z-w)^4}+\frac{2T(w)}{(z-w)^2}+\frac{\partial
T(w)}{(z-w)},\eea with central charge $c=1=\frac{k\,\times {\rm sdim}\left(D(2,1;\a)\right)}{k+0}$. Moreover, it is easy
to check that
\bea
T(z)X(w)=\frac{X(w)}{(z-w)^2}+\frac{\partial X(w)}{z-w}
        =\partial_{w}\left\{\frac{X(w)}{z-w}\right\},\quad\quad\forall X\in D(2,1;\a).
\eea Namely, all
the currents are primary fields of conformal dimension one


\section{Screening currents}
\label{SC}  \setcounter{equation}{0}

An important object in applying the free field realization to the
computation of correlation functions  of the associated CFT is
screening currents. A screening current is a primary field with
conformal dimension one and has the property that the singular
part of the OPE with the affine currents is a total derivative.
These properties ensure that integrated screening currents
(screening charges) may be inserted into correlators while the
conformal or affine Ward identities remain intact. This in turn
makes them very useful in computation of the correlation functions
\cite{Dos84,Ber90}. For the present case, we find the following three screening
currents associated with the simple roots:
\bea
S_{\a_1}(z)&=&\left\{\g_{\a_3}(z)\g_{\a_2}(z)\Psi_{\a_1+\a_2+\a_3}(z)
              +(1+\a)\g_{\a_2}(z)\Psi^+_{\a_1+\a_3}(z)\b_{2\a_1+\a_2+\a_3}(z)\right.\no\\
           &&\quad\quad+(1+\a)\Psi^+_{\a_1+\a_2+\a_3}(z)\b_{2\a_1+\a_2+\a_3}(z)
             -\g_{\a_3}(z)\Psi_{\a_1+\a_3}(z)\no\\[6pt]
           &&\quad\quad-\left.\g_{\a_2}(z)\Psi_{\a_1+\a_2}(z)+\Psi_{\a_1}(z)
              \right\}\,e^{\sqrt{\frac{1+\a}{2k}}\phi_1(z)+\sqrt{\frac{1}{2k}}\phi_2(z)
              \sqrt{\frac{\a}{2k}}\phi_3(z)},\label{Screening-1}\\[6pt]
S_{\a_2}(z)&=&\b_{\a_2}(z)\,e^{-\sqrt{\frac{2}{k}}\phi_2(z)},\\[6pt]
S_{\a_3}(z)&=&\b_{\a_3}(z)\,e^{-\sqrt{\frac{2\a}{k}}\phi_3(z)}.
\label{Screening-2}\eea The normal ordering of the free field
expressions is implicit in the above equations. The
OPEs of the screening currents with the energy-momentum tensor (\ref{Tensor}) and
the $D(2,1;\a)$ currents (\ref{Free-field-realization-1})-(\ref{Free-field-realization-2}) are
\bea
&& T(z)\,S_{\a_j}(w)=\frac{S_{\a_j}(w)}{(z-w)^2}+\frac{\partial S_{\a_j}(w)}{(z-w)}
=\partial_w\lt\{\frac{S_{\a_j}(w)}{(z-w)}\rt\},\,\,j=1,2,3,\\[6pt]
&&E_{\a_i}(z)\,S_{\a_j}(w)=0,\quad\quad H_{\a_i}(z)S_{\a_j}(w)=0,
  \quad\quad i,j=1,2,3,\\[6pt]
&&F_{\a_1}(z)\,S_{\a_j}(w)=\d_{1j}\,\partial_w\left\{
   \frac{k\,e^{\sqrt{\frac{1+\a}{2k}}\phi_1(w)+\sqrt{\frac{1}{2k}}\phi_2(w)+\sqrt{\frac{\a}{2k}}\phi_3(w)}}
   {z-w}\right\},\quad\quad j=1,2,3,\\[6pt]
&&F_{\a_2}(z)\,S_{\a_j}(w)=\d_{2j}\,\partial_w\left\{
   \frac{k\,e^{-\sqrt{\frac{2}{k}}\phi_2(w)}}
   {z-w}\right\},\quad\quad j=1,2,3,\\[6pt]
&&F_{\a_3}(z)\,S_{\a_j}(w)=\d_{3j}\,\partial_w\left\{
   \frac{\frac{k}{\a}\,e^{-\sqrt{\frac{2\a}{k}}\phi_3(w)}}
   {z-w}\right\},\quad\quad j=1,2,3.
\eea The screening currents obtained here  correspond to the
screening currents of the first kind \cite{Ber86}. The screening current $S_{\a_1}(z)$ is fermionic
and the others are bosonic.


\section{Discussions}
\label{Con} \setcounter{equation}{0}

We have studied the exceptional current algebra
$\widehat{D(2,1;\a)}$  at general level
$k$. We have constructed the Wakimoto free field realization of the currents
(\ref{Free-field-realization-1})-(\ref{Free-field-realization-2}) and the
energy-momentum tensor (\ref{Tensor}).
We have also found three screening currents,
(\ref{Screening-1})-(\ref{Screening-2}), of the first kind. With a special choice
of $\a=1$, the results recover those of $osp(4|2)$ current algebra \cite{Yan09}.

\section*{Acknowledgements}
The financial supports from  the National Natural Science
Foundation of China (Grant Nos. 10931006, 10975180, 11075126, 11031005 and 11047179) and
Australian Research Council  are gratefully acknowledged.
WLY would like to thank the School of Mathematics and Physics, The University of Queensland,
where part of this work was done,  for
hospitality and support through a Raybould Visiting Fellowship.

\vspace{1.00truecm}



\begin{thebibliography}{99}
\bibitem{Ber99} N. Berkovits, C. Vafa and E. Witten, {\it JHEP\/},
{\bf 03} (1999), 018.
\bibitem{Bers99} M. Bershadsky, S. Zhukov and A. Vaintrob, {\it
Nucl. Phys.\/} {\bf B 559} (1999), 205.
\bibitem{Sch06} V. Schomerus and H. Saleur, {\it Nucl. Phys.\/}
{\bf B 734} (2006), 221; {\it Nucl. Phys.\/} {\bf B 775} (2007),312.
\bibitem{Que07} T. Quella and V. Schomerus, {\it JHEP} {\bf 09} (2007), 085.
\bibitem{Efe83} K. Efetov, {\it Adv. Phys.\/} {\bf 32}
      (1983), 53.
\bibitem{Ber95}D. Bernard, {\tt hep-th/9509137}.
\bibitem{Mud96} C. Mudry, C. Chamon and X.\,-G. Wen,
      {\it Nucl. Phys.\/} {\bf B 466} (1996), 383.
\bibitem{Maa97} Z. Maassarani and D. Serban, {\it Nucl. Phys.\/}
      {\bf B 489} (1997), 603.
\bibitem{Bas00} Z.\,S. Bassi and A. LeClair, {\it Nucl. Phys.\/}
      {\bf B 578} (2000), 577.
\bibitem{Gur00} S. Guruswamy, A. LeClair and A.\,W.\,W. Ludwig,
      {\it Nucl. Phys.\/} {\bf B 583} (2000), 475.
\bibitem{Lud00} A.\,W.\,W. Ludwig, {\tt cond-mat/0012189}.
\bibitem{Bha01} M.\,J. Bhaseen, J.\,-S. Caux, I.\,I. Kogan and
     A.\,M. Tsveilk, {\it Nucl. Phys.\/} {\bf B 618} (2001), 465.
\bibitem{Mat08} T. Matsumoto and S. Moriyama, {\it JHEP} {\bf 04} (2008), 022;
{\it JHEP} {\bf 09} (2009), 097.
\bibitem{Fra97} P. Di Francesco, P. Mathieu and D. Senehal, {\it
Conformal Field Theory\/}, Springer Press, Berlin, 1997.
\bibitem{Fei90} B. Feigin and E. Frenkel, {\it Commun. Math.
Phys.\/} {\bf 128} (1990), 161.
\bibitem{Bou90} P. Bouwknegt, J. McCarthy and K. Pilch, {\it Prog.
Phys. Suppl.\/} {\bf  102} (1990), 67.
\bibitem{Ito90} K. Ito, {\it Phys. Lett.\/} {\bf B 252} (1990), 69.
\bibitem{Fre94} E. Frenkel, {\tt hep-th/9408109}.
\bibitem{Ras98} J. Rasmussen, {\it Nucl. Phys.\/} {\bf B 510} (1998), 688.
\bibitem{Din03} X.\,-M. Ding, M. Gould and Y.\,-Z. Zhang, {\it
Phys. Lett.\/} {\bf A 318} (2003), 354.
\bibitem{Zha05} Y.\,-Z. Zhang, X. Liu and W.\,-L. Yang, {\it Nucl.
Phys.\/} {\bf B 704} (2005), 510.
\bibitem{Yan06} W.\,-L. Yang, Y.\,-Z. Zhang and X. Liu, {\it Phys. Lett.}
{\bf B 641} (2006), 329; {\it J. Math. Phys.\/} {\bf 48} (2007), 053514.
\bibitem{Yan08} W.\,-L. Yang and Y.\,-Z. Zhang, {\it Phys. Rev. \/} {\bf D 78}
(2008), 106004.
\bibitem{Yan09} W.\,-L. Yang, Y.\,-Z. Zhang and S. Kault, {\it Nucl.
Phys.\/} {\bf B 823} (2009),  372.
\bibitem{Fra96} L. Frappat, P. Sorba and A. Scarrino,
Dictionary on Lie algebras and superalgebras, Academic Press,
     New York, 2000.
\bibitem{Dos84} VI.\,S. Dotsenko and V.\,A. Fateev, {\it Nucl.
Phys.\/} {\bf B 240} (1984), 312; {\it Nucl. Phys.\/} {\bf B 251} (1985), 3691.
\bibitem{Ber90} D. Bernard and G. Felder, {\it Commun. Math.
Phys.\/} {\bf 127} (1990), 145.
\bibitem{Ber86} M. Bershsdsky and H. Ooguri, {\it Commun. Math.
Phys.\/} {\bf 126} (1986), 49.


\end{thebibliography}
\end{document}